\documentclass{article}
\usepackage{amssymb}
\usepackage{amsmath}
\usepackage{graphicx}

\usepackage{wrapfig}
\usepackage{array}
\usepackage{bm}
\usepackage{subfigure}
\usepackage [english]{babel}
\usepackage[dvips]{epsfig}
\usepackage{slashbox}
\newcommand{\lb}{\label}
\newcommand{\be}{\begin{equation}}
\newcommand{\ee}{\end{equation}}
\newcommand{\ba}{\begin{align}}
\newcommand{\ea}{\end{align}}
\newcommand{\bea}{\begin{eqnarray}}
\newcommand{\eea}{\end{eqnarray}}
\newcommand{\bw}{\begin{widetext}}
\newcommand{\ew}{\end{widetext}}

\newcommand{\e}{{\rm e}}

\begin{document}

\title{Transplanckian radiation
in theories with extra dimensions}
\author{D.V. Gal'tsov\\ {\small  Department of Theoretical Physics, Moscow State University,
119899, Moscow, Russia} \\{\small {\it Email} :
galtsov@phys.msu.ru}}

\date{}
\maketitle
\begin{abstract}
We discuss whether particles undergoing transplanckian collisions
in TeV-scale gravity can deplete most of their energy on bremsstrahlung
for impact parameters much larger than the gravitational radius
of the presumably created black hole.
\end{abstract}
\section{Introduction}

The models of TeV-scale Quantum Gravity (TQG)  proposed as an
attempt to solve the hierarchy problem
open the possibility to study  physics beyond the Planck's scale. A
conceptually and technically simplest TQG model suggested by
Arkani-Hamed, Dimopoulos and Dvali (ADD) \cite{ADD} and further elaborated in
\cite{GRW} assumes that
the standard model  particles reside in the four-dimensional
sector of space-time ${\mathcal M}_4$, while gravity propagates in the $D$-dimensional
bulk with the $d=D-4$ flat dimensions compactified on a torus
${\mathcal T}^d$.
$D$-dimensional gravity is strong, the corresponding Planck mass
$M_{*}$ being of the order of  TeV.   The Planck length $l_*$ and
the gravitational radius $r_s$ associated with the center-of mass
energy $\sqrt{s}$ in the  ADD model are  \vspace{-.1cm}
 \be\label{rs}
 l_{*}=\left(\frac{\hbar G_D}{ c^3}\right)^{\frac1{d+2}}\sim
 \frac{\hbar}{M_{*}c},  \;\;\;\;  r_s=\left(\frac{G_D
 \sqrt{s}}{c^4}\right)^{\frac1{d+1}}.
 \ee
   In the transplanckian (TP) particle collisions gravity not only becomes the dominant
force, but it partly restores its
classical character \cite{GiRaWeTrans}.
Since the gravitational radius $r_s$ corresponding to the energy
in the center of mass frame  grows with $s$,
while the  de Broglie length of the
particles $  \lambda_{B}=\hbar c/\sqrt{s}$ decreases, therefore for
$\sqrt{s}>M_*$  the classicality condition is satisfied \vspace{-.1cm}
 \be\lambda_{B}\ll l_*\ll r_S. \vspace{-.1cm}
 \ee
In the ADD model with $d\neq 1$ the classical TP region is also restricted
from above ~\cite{GiRaWeTrans} by the impact parameter \vspace{-.1cm}
 \be
 \label{bece}
 b_c\equiv \frac1{\sqrt{\pi}}\left( \frac{\varkappa_D^2\Gamma(d/2)\,s}
 {16\pi}\right)^{1/d}.\vspace{-.1cm}
 \ee
We have therefore the following three TP sectors:
\cite{Giddings:2007qq}

1) $b>b_c.\;$ This is the quantum sector where the scattering is
dominated by the one-graviton exchange.

2) $r_s<b<b_c.\;$ This is the eikonal sector, where the scattering amplitude is
dominated by the sum of ladder and cross-ladder diagrams whose
summation gives the eikonal amplitude \cite{Kabat:1992tb}. The
stationary phase approximation of the latter coincides with the
classical amplitude \cite{Gal'tsov:2009mj}. It is also known that
the semiclassical calculation of TP elastic scattering cross
section in four-dimensional space-time \cite{tHooft} agrees with the
string theory result \cite{acv2}. Gravitational radiation in this
region is expected to be well-described by the classical theory.

3) $b<r_s.\;$ This is the region of strong gravity, where the
main process is  the formation of the black hole
\cite{BH}. This was  checked within an  approach based on the picture
of colliding waves representing
ultrarelativistic particles \cite{EaGi}. However
this approximation is susceptible to radiation
reaction \cite{Gal'tsov:2009zi}, which is still not well understood.
Various more sophisticated approaches to test the
conjecture of black hole creation at colliders were suggested
(see a recent review
\cite{Park:2012fe}), which gave additional arguments of its
validity. Numerical work  performed in this direction
is reviewed in  \cite{Cardoso:2012qm}).

Here we discuss whether gravitational radiation in TP collisions can be large enough
in the region 2). If so, this could substantially modify
the predicted cross-section of the formation of a black hole.
Note, that for  a {\em head-on} collision of black holes the upper bound was given by
Eardley and Giddings \cite{EaGi} generalizing Penrose limit: \vspace{-.2cm}
\be
\epsilon\leq 1 -\frac12\left(\frac{(D-2)\Omega_{D-2}}{2\Omega_{D-3}}\right)^{\frac1{D-2}},
 \vspace{-.1cm}\ee
which gives the bound about $41,9\% $ for $D=11$.
A recent calculation \cite{Herdeiro:2011ck} using  the  approach of D'Eath and Payne
\cite{D'Eath:1992hb} gave in the first perturbative order the result slightly
below this bound (the second order and numerical calculations
give smaller values).
However, it is known that radiation  from particles
plunging into the black hole {\em grow with the impact parameter}. Moreover,
the above calculations are based on the picture of colliding waves,
which itself is appropriate only if radiation losses are small,
while the above values can not be considered small.

\section{Eikonal and bremsstrahlung}
The eikonalized elastic scattering amplitude can be presented as the
integral \vspace{-.1cm} \be {\cal M}_{\rm
eik}(s,t)=2is\int\e^{i{\bf q\cdot
b}}\left(1-\e^{i\chi(s,b)}\right)d^2b, \label{eik} \vspace{-.1cm}
 \ee
where the two-dimensional vectors ${\bf q,\,b}$ lie in the
transverse plane, with $\bf b$ playing the role of the impact parameter
vector. The
transverse component $\bf q$ of the momentum transfer in this
approximation satisfies ${\bf q}^2 \approx -q^\mu q_\mu$, so that
$t\simeq -{\bf q}^2$. This expression in the usual four-dimensional
theory corresponds to summation of the ladder and crossed-ladder
diagrams. The eikonal phase $\chi(s,b)$ can be obtained
expanding the exponential and equating the leading term
to the Born amplitude \vspace{-.1cm}\be
 \chi(s,b) =\frac1{2s}\int\e^{-i{\bf q\cdot b}} {\cal M}_{\rm Born}(s,t)\; \frac{d^2
 q}{(2\pi)^2}
 \label{eikinvers}. \vspace{-.1cm}
 \ee
The computation gives%\vspace{-.1cm}
\be
 \chi(s,b)=\left(\frac{b_c}{b}\right)^d,
 \label{eikexpli} %\vspace{-.1cm}
  \ee
where $b_c$ is given by (\ref{bece}). Substituting this into (\ref{eik})
and calculating the integral in the stationary phase
approximation around the point \vspace{-.2cm}\be b_s =\left(\frac{d
 b_c^d}{q}\right)^{1/(d+1)},\vspace{-.2cm}\ee
one recovers the purely classical result:
 \vspace{-.1cm}
\be
 {\cal M}_{\rm
 cl}(s,t)= \frac{4\sqrt{\pi} s\e^{i(qb_s-\pi/2)}}{q\sqrt{d+1}}
  \left( \frac{2 \sqrt{\pi} s \Gamma(d/2+1)}
 {M_*^{d+2} q } \right)^{ \frac{1}{d+1}},
 \label{stat}\vspace{-.1cm}
\ee which was obtained classically for the small angle scattering of
ultrarelativistic point particles
\cite{Gal'tsov:2009mj}.

Now consider gravitational bremsstrahlung, which is the main
inelastic process in the eikonal region of TP scattering. In the
eikonal approach  radiation manifests itself as the imaginary part
$\chi(s,b)$ which can be found taking into account the leading corrections to
the ladder approximation from the called H-diagrams \cite{ACV}. According to
\cite{GiRaWeTrans}, \vspace{-.2cm}\be {\rm Im} \chi\sim
\left(\frac{b_r}{b}\right)^{3d+2}, \vspace{-.1cm}
 \ee
where one more length parameter $b_r$ is introduced, satisfying the
relation\vspace{-.2cm}
 \be
  \frac{b_r}{r_s}=\left(\frac{b_c}{r_s}\right)^{\frac{d}{3d+2}}.\vspace{-.1cm}
 \ee
 In the situation when the eikonal sector of TP scattering is
wide enough, $r_s\ll b_c$, one has $b_r\gg r_s$ and the impact parameter
 $b_r$ can lie in
the interval \vspace{-.2cm}\be r_s\ll b \ll b_r, \vspace{-.2cm}\ee in which case ${\rm Im} \chi$ is not
small. If the imaginary part of the eikonal is interpreted as the
number of emitted gravitons whose frequency
is much higher than $\omega_b= 1/b$, this would lead to the conclusion that
the colliding particles can deplete all the energy for impact
parameters much larger than $r_s$.   Meanwhile,  extraction of the eikonal
imaginary part  from the H-diagram probably is only consistent
in the deep infrared region.  Assuming that the spectrum is
dominated by the frequencies $\omega<\omega_b$, and the number  of
soft gravitons is   $N={\rm Im} \chi$, one finds the following
estimate for the radiation efficiency \cite{GiRaWeTrans} \be
\epsilon=\frac{\Delta E}{E}\sim
\left(\frac{r_s}{b}\right)^{\frac{d}{3d+3}}, \ee which is not
catastrophical for $b> r_s$. However,
classical considerations \cite{Gal'tsov:2010me} indicate
that the dominant region
of the bremsstrahlung spectrum is $\omega\gg\omega_b$. Thus, one needs other methods
to answer the question whether $\epsilon$ may become of the order of unity
for $b>  r_s$. An interesting approach to TP bremsstrahlung problem
using the eikonal approximation in the spirit of 't Hooft was suggested by Lodone
and Rychkov \cite{Lodone:2009qe} , but so far it was applied only to gluons.

\section{TP bremsstrahlung at $b\gg r_s$}
In a series of papers
\cite{Gal'tsov:2009zi,Gal'tsov:2010me,Constantinou:2011ju}
we calculated ultrarelativistic bremsstrahlung in ${\mathcal M}_4\times {\mathcal T}^d$
using classical perturbation theory
in momentum space, which
was suggested long ago in the context of four-dimensional General
Relativity \cite{GGM}. Starting with the action \be \lb{1}
S=-\sum\int m\sqrt{g_{MN}{\dot x}^M {\dot x}^N } ds
+S_{\phi}(\phi, g)+S_{\rm int}\left[x(s),\phi,g\right]+S_g(g), \ee for two
point particles, mutually interacting with the set of  non-gravitational
fields $\phi$ and the gravity described by
bulk metric $g_{MN}$. The metric is presented as $g_{MN}=
\eta_{MN}+\kappa_D h_{MN}$, and $x(s),\; \phi$ and $h_{MN}$ are
further expanded in terms of the particle-field couplings $f$ and the
gravitational coupling $\kappa_D$.  In the zero order approximation
particles move freely in the opposite directions at an impact parameter $b$.
Then we iterate the system of the particle equations of motion
and the field equations up to the second order, in which radiation is manifest.
This procedure  presumably converges in the ultrarelativistic case when the
scattering angle $\theta'$ is small, though to get more precise limits of applicability
on has to go to the next iteration order, which is quite non-trivial.

One is interested
in computing the total radiation efficiency $\epsilon$ and the spectral distribution
under different assumptions about nature of the field dominating the
interaction between the particles (mediator field), and nature of the radiation field.
The set of $\phi$ generically contains the brane $\varphi$ and the bulk $\Phi$
fields either of which can be mediator and/or radiated field. Gravity  interacts
with particles and with both fields $\varphi,\;\Phi$, introducing non-linearity into
the problem.
Coupling constants of the particles with $\varphi\;(f_0)$ and  $\Phi\;(f_d)$
have different dimensions and are
related though the volume of the torus $V_d$ as $f_d^2=f_0^2 V_d$. The corresponding classical
length parameters $r_d$ are introduced via the relation $f_d^2/r_d^{d+1}=m$,
so $r_d^{d+1}=r_0^2 V_d$.
The bulk fields $\Phi,\;h_{MN}$ depending on $x^M=(x^\mu,\;y^i)$
are expanded in the Kaluza-Klein modes $\Phi_n(x),\;h_{MN}^n(x),\;n\in Z^d$  which
have the masses $\mu_n^2=(2\pi {\bf n}^2/L^2)$.

The radiation efficiency and its
spectrum  depend on the nature of $\phi$ and on whether gravity
is  the dominant mediator. If not,
one deals with  the flat space problem in which both mediating and emitted
fields are  linear. This is
the case of the Maxwell-Lorentz theory, which was exhaustively explored
both in classical and quantum electrodynamics. The only novel here feature is presence of
extra dimensions. Note that if one deals only with bulk fields, the ADD problem in
${\mathcal M}_4\times {\mathcal T}^d$  reduces to that in $D$-dimensional Minkowski
space provided the large number of KK modes in involved.

The frequency of radiation depends on the emission angle with
respect to the direction of the collision
$
 \omega_{\rm cr}(\theta)\simeq  b^{-1}(\theta^2+\gamma^{-2})^{-1} ,
 $
where $\gamma$ is the Lorentz factor in the rest frame
of one of the particles. So most of the
radiation is beamed inside the cone
$
\theta < 1/{\gamma}
$
in any dimension $D$. The maximal frequency of radiation is \vspace{-.2cm}
 \begin{equation}
 \omega_{\rm max}\simeq \omega_{\rm cr}(\theta=0) \sim
\frac{2\gamma^2}{b},
\end{equation}
whose vicinity gives the dominant contribution.
The total bremsstrahlung efficiency for different combinations
of exchange modes and radiated modes reads \cite{Gal'tsov:2010me}
(omitting numerical coefficients):
\begin{enumerate}
\item
brane modes exchanged and radiated:
$\epsilon_{\phi\phi} \sim    \gamma \, r_0^3 / b^3$;
\item
bulk modes exchanged  brane modes radiated:
$\epsilon_{\Phi\phi}\sim
  \gamma \, \frac{r_0}{b}\, (\frac{r_d}{b})^{2(1+d)}$;
\item brane modes exchanged, bulk modes radiated:
$\epsilon_{\phi\Phi}
\simeq     \left(\frac{r_0}{b}\right)^2
\left(\gamma\,\frac{r_d}{b}\right)^{1+d}$;
\item
bulk modes exchanged and radiated:
$\epsilon_{\Phi\Phi}
\simeq  (\gamma\, r_d^3 / b^3 )^{1+d}$.
\end{enumerate}
These formulas can be interpreted in terms of the effective numbers of massive KK states
contributing as exchange modes
$N_{\rm ex}\sim V_d/b^d$ and radiated modes $N_{\rm rad}\sim V_d\gamma^d/b^d$,
where the $\gamma-$enhancement factor in $N_{\rm rad}$ accounts for modes
with masses up to $\omega_{\rm max}$ emitted inside the cone
$
\theta_n < {1}/{\gamma}.
$
in the bulk. Extra dimensions contribute to radiation efficiency with the factor
$N_{\rm ex}^2$ for KK exchange  modes and with the factor
$N_{\rm rad}$ for radiation modes, so, e.g.,
$\epsilon_{\Phi\Phi}\sim \epsilon_{\phi\phi} N_{\rm ex}^2N_{\rm rad}$.
This explains the origin of an extra factor $\gamma^d$ in $\epsilon_{\Phi\Phi}$.
Therefore, classically, the bremsstrahlung efficiency grows with $\gamma$,
and the $\gamma$-factor is $d$-dependent.

However, one has the quantum restriction on the frequency $\hbar\omega<m\gamma$,
which for the maximal frequency $\omega=\omega_{\rm max}=2\gamma^2/b$  gives
$b_{\rm min}=\lambda_c\gamma $, where
$\lambda_c=1/m $ is the Compton length.
The boundary value of the radiation efficiency $\epsilon_{\Phi\Phi}
\simeq  ( r_d^3 / (\lambda_c^3 \gamma^2))^{1+d}$ thus decreases with energy.

The situation becomes more complicated if gravity is the dominant
mediator \cite{Constantinou:2011ju}.
The main new feature is non-linearity of the problem due to $\phi\phi h$
vertex and the three-graviton  coupling, which leads to non-locality of the source in the
D'Alembert equation for radiation modes:
\begin{equation}
\Box \Phi_n^{\rm rad}=j_n\equiv\rho_n+\sigma_n.
\end{equation}
Here $\rho_n$ has support on the particles world-lines, while $\sigma_n$ is given
in terms of lower order field perturbations; it is extended in space
(including extra dimensions). It turns out that $\rho_n$ and $\sigma_n$
compete with each other and the result depends on the number
of extra dimensions. In the case $d=0$,  as it was shown long ago \cite{GGM},
these contributions mutually cancel in the frequency range
\begin{equation}\label{hf}
\omega'_{\rm max}<\omega<\omega_{\rm max},\qquad
 \omega'_{\rm max}=\gamma/b=\omega_{\rm max}/2\gamma,
\end{equation}
so the frequency is bounded by $\omega'_{\rm max}=\gamma/b$.
In this case the quantum boundary $\omega'_{\rm max}<m \gamma$
is energy-independent.
The efficiency of the gravitational bremsstrahlung in the gravity
mediated collision in $D=4$ \cite{GGM} is
\begin{equation}\epsilon_{hh}
\sim     \gamma_{\rm cm}\, r_s^3  / b^3,  \end{equation}
where $r_s$ is given by (\ref{rs}) with $d=0$.
The energy-dependent restriction on $b$ arises from
the condition of smallness of the gravitational potential
energy of the fast particle in the rest frame of the other,
with respect to the particle energy, equivalent to
$b\gg r_s\gamma_{\rm cm}$.
Thus, on the boundary $\epsilon_{hh}$ remains small.

For $d\neq 0$ cancelation of local and non-local contributions
also takes place but less complete,
so the spectral-angular distribution still contains higher
frequencies (\ref{hf}). The total efficiency in $D=5$ scales as
$\epsilon_{hh}\sim\gamma_{\rm cm}\,\ln\gamma_{\rm cm}\cdot r_s^6  / b^6  $,
while for $d\geq 2$ one obtains
\begin{equation}\epsilon_{hh}
\sim    \left(r_s  / b \right)^{3(d+1)} \gamma_{\rm cm}^{2d-1}.
\end{equation} Smallness of the scattering angle with respect to
$\gamma^{-1}$ (this restriction generalizes the one relevant for
small angle bremsstrahlung calculation on the fixed background, in
our two-body approach it may be overrestricting) implies $b>
r_s\gamma_{\rm cm}^{1/d+1}$, so classically we get on the boundary
$\epsilon_{hh} \sim      \gamma_{\rm cm}^{2(d-2)}.  $ This means
that for  $d\geq 3$ the energy is depleted for $b\gg r_s$. This
result is still susceptible to quantum bounds, since for $d\neq 0$
the quantum frequency restriction on $b$ is stronger than classical.
This question is currently is under investigation
\cite{Gal'tsov:2012sn}. If confirmed, the above estimate means that
classical radiation damping may be regarded as another
classicalization mechanism additional to creation of the black hole.

\section*{Acknowledgments}
The author thanks the Organizing Committee of ICGA-12 for an excellent conference.
He is grateful to G.Kofinas, P.Spirin and Th.Tomaras for useful
comments and fruitful collaboration. The work was supported by
the RFBR grant 11-02-01371-a.
%\vspace{0.5cm}

\end{document}